\documentclass{eptcs}
\usepackage{breakurl}   
\usepackage{amssymb}
\usepackage{graphicx}
\usepackage{multirow}
\usepackage{amsmath}
\usepackage{tikz}
\usepackage{latexsym}
\usepackage{amsfonts}
\usepackage{graphics}
\usepackage{stackrel}
\usepackage{hyperref}
\usepackage{tikz}
\usepackage{alltt}
\usepackage{array}
\usepackage{times}
\usetikzlibrary{arrows}
\usetikzlibrary{matrix}
\usepackage{mathtools,bussproofs}

\title{MELA: Modelling in Ecology\\ with Location Attributes}
\author{Ludovica Luisa Vissat$^1$, Jane Hillston$^1$, Glenn Marion$^2$, Matthew J. Smith$^3$\\
\institute{$^1$ LFCS, School of Informatics, University of Edinburgh, UK\\
$^2$ Biomathematics and Statistics Scotland, Edinburgh, UK\\
$^3$ Computational Science Laboratory, Microsoft Research, Cambridge, UK}}
\def\titlerunning{MELA: Modelling in Ecology with Location Attributes}
\def\authorrunning{L. Luisa Vissat \& J. Hillston \& G. Marion \& M. J. Smith}
\begin{document}
\maketitle

\begin{abstract}
Ecology studies the interactions between individuals, species and the environment. The ability to predict the dynamics of ecological systems would support the design and monitoring of control strategies and  would help to address pressing global environmental issues. It is also important to plan for efficient use of natural resources and maintenance of critical ecosystem services. The mathematical modelling of ecological systems often includes nontrivial specifications of processes that influence the birth, death, development and movement of individuals in the environment, that take into account both biotic and abiotic interactions. To assist in the specification of such models, we introduce MELA, a process algebra for Modelling in Ecology with Location Attributes.  Process algebras allow the modeller to describe concurrent systems in a high-level language. A key feature of concurrent systems is that they are composed of agents that can progress simultaneously but also interact --- a good match to ecological systems.  MELA aims to provide ecologists with a straightforward yet flexible tool for modelling ecological systems, with particular emphasis on the description of space and the environment.  Here we present four example MELA models, illustrating the different spatial arrangements which can be accommodated and demonstrating the use of MELA in epidemiological and predator-prey scenarios. 
\end{abstract}

\section{Introduction}

In this work we present a new process algebra called MELA, developed for Modelling in Ecology with Location Attributes. It is suitable for formally describing ecological systems, with focus on space abstraction and environmental description. Process algebras were introduced in concurrency theory, to model systems where processes may interact with each other and several actions may be executed simultaneously. In recent years there has been considerable interest in applying these formal languages and their benefits in fields other than computer science; particular examples include biology \cite{bio-pepa}, public transportation systems \cite{cheng} and collective adaptive systems \cite{carma}. Defining a process algebra for application in ecology is worthwhile for many different reasons. First of all, process algebras are simple high-level languages, that give the possibility to easily model and to better understand the interactions between agents and whole system properties, for example, consistency check via static analysis. Another important aspect of process algebras is their compositional nature: the \textit{individual-based} approach of modelling populations enables the modeller to describe the evolution of each individual in the population as a process and, subsequently, to compose a set of individuals into a complete ecological system \cite{palps}. Therefore, MELA allows users to describe single or multiple species (i.e.\ community) population models. Furthermore, it incorporates a structural operational semantics: this provides an interpretation of the expressions as a labelled transition system, which will support the implementation of the language and simulation of models. This paper presents what we believe will be the foundations of further contributions in the area of formal modelling in ecology, for which MELA and its associated software tool will be the vehicle.  Like other stochastic process algebras MELA will support complementary analysis techniques, such as stochastic simulations, numerical solution of ODEs (ordinary differential equations), and analyses based directly on the underlying  continuous time Markov chain (CTMC). Ecologists have recognised that individual-based Markov chains models present a useful alternative method to the classic population-based ODEs to describe dynamics of a system  \cite{primer}. These stochastic models are more suitable in systems that present variability and small interacting populations. Furthermore, if space is considered, the variability increases and making the use of stochastic models even more appropriate.

Therefore, in MELA, models are stochastic and individual-based, as well as spatially-explicit. Spatial heterogeneity has been recognized as of the key importance in ecology \cite{spatial}. Accordingly, MELA takes account of the spatial description of the system, providing location attributes for the agents described in MELA models. The importance of models is well-known in ecology \cite{fmd}, where we can use them to predict the system behaviour and use this predictive power to support decision-making. For example, in the case of the spread of diseases, models are used to find the optimal control strategies for containing the spread, in terms of time and impact, or to predict the results of a population reduction, as a key way to control disease in livestock and wildlife, as in \cite{popred}. Models are also widely used to plan for efficient use of resources (e.g.\ foraging \cite{foraging}), predicting the evolution of the system and facilitating management process. Although some process algebras have already been used for ecological modelling \cite{bio-pepa}, \cite{palps}, these existing languages were developed for other purposes (e.g.\ Bio-PEPA for biochemistry) and are lacking in support for key aspects of ecological modelling. For example, they do not provide explicit support for spatial structure, making exploration of the impact of different structures cumbersome and laborious. More fundamentally, this limits tools for analysis to just temporal properties, not directly taking into account spatial aspects. In order to inform the creation of the new language, we analysed past works to identify important features, but also missing ones that we have to bear in mind. Therefore, we created MELA, a process algebra with the objective to be useful for ecologists; thus we set out to design a language that is both compact and user-friendly. We have a separate and clear description for the spatial structure and the agents' behaviour. The modeller can choose between different spatial structures (graph, discrete line segment, 2D grid, 3D grid, nested spatial structure) without dealing directly with the set of locations and the population vector, that will be created automatically. Moreover, the separation and the high-level description allows the modeller to better comprehend the model itself and to easily make changes. In particular these features are intended to assist the modeller in understanding the effect of changes in spatial structure on the dynamics of the system, by making it easy to consider the same agents in different spatial environments. MELA will support different types of analysis of ecological systems but our main focus is on spatio-temporal properties, to better understand, as mentioned above, how the spatial structure influences the dynamics of the system (such as dispersal, spread of diseases). In addition, MELA features will help to improve the model checking techniques, using the ability to tag and distinguish the different type of actions that agents can undertake (e.g.\ demographic changes, spatial movement) as well as keeping track and counting the number of the locations where a specific action happens, allowing, for example, the study of the movement of agents across space.

The paper is structured as follows. In the next section we present three existing process algebras and discuss their features, related to their application in ecology. Section 3 describes the syntax of MELA and, in Section 4, we define MELA semantics, in terms of a labelled transition system. Section 5 presents four examples, in which we show a grid spatial abstraction in a susceptible/infective (SI) model, a graph spatial abstraction in a prey-predator model, the presence of environment factors in the spread of a Cholera epidemic and the coexistence of nested locations, in this case a grid/graph spatial abstraction, to model local/global interactions in the spread of a disease between farms. Finally, Section 6 reports discussion and future directions for investigation.

\section{Current formalisms for ecological modelling}

In this section we review the suitability of three existing process algebras that have been used for, if not designed for, ecological purposes.
If we want to model an ecological system, it is natural to consider agents
that interact with each other and with the environment, as well as making
actions which may depend on some aspects of the system. We want to
describe the space where the agents live and we have also to consider variations in the number of individuals of each species, taking into account
processes like birth, death and migration.
Agents can interact with others that are in their neighbourhood or even
with more distant agents, exchanging messages, and making
decisions based on these interactions, depending on their personal needs and characteristics.
Therefore, key requirements for a process algebra to model ecological interactions are modelling \textit{birth-death} processes, an ability to describe \textit{space} and concepts of \textit{neighbourhood}, the description of \textit{actions} and the \textit{influence} they may have on
the behaviour of the agents. We will deal with each of these features,
starting from the past work, showing important and missing aspects from the ecological point of view.\\
The pioneers of using process algebras for ecological modelling were the group of Norman and Shankland in the Department of Computing Science and Mathematics of Stirling University with the project \textit{System Dynamics from Individual Interactions: A process algebra approach to epidemiology} \cite{stirling}. They were the first to underline the importance and usefulness of process algebra as a modelling methodology for epidemiology. Their work makes use of the Weighted Synchronous Calculus of Communicating Systems (WSCCS) \cite{wsccs}, a discrete time process algebra with weighted choice and no spatial information, and of Performance Evaluation Process Algebra (PEPA) moving to actions occurring in continuous time at a specified rate and choice driven by competition between rates \cite{pepastirling}. However, this first approach does not consider spatial heterogeneity, a fundamental requirement in ecological modelling, therefore here we focused on three other process algebras with this feature.\\
Bio-PEPA \cite{original} is a process algebra developed for the modelling and analysis of biochemical networks, and it is a modification of PEPA (Performance Evaluation Process Algebra) \cite{pepa}, a process algebra originally developed for modelling computer and communication systems. A variant of Bio-PEPA is presented in \cite{bio-pepa}, adapted for epidemiological modelling. Bio-PEPA presents an easy and straightforward way to describe birth/death processes while, due to the biochemical background, it is not possible to directly describe movement of the agents in the system, an important aspect for ecological modelling.\\
PALPS (Process Algebra with Locations for Population Systems) \cite{palps} is described as ``the first process-algebraic framework developed specifically for reasoning about ecological models as well as in its treatment of a state and its capability of expressing state-dependent behaviour". This process algebra can be considered to be as an extension of CCS \cite{ccs} with probabilistic choice, locations and location attributes. The ability to define location-dependent behaviour is an important requirement for ecological modelling but there are still some weaknesses in this formalism.
In PALPS we have to describe the behaviour and list all the agents, explicitly defining each agent with its location attribute, while with MELA the set of possible locations $\mathcal{L}$ is defined at the beginning. Therefore, the different agents are described just once, not for each location but writing them with a location attribute $l$. Moreover, there is no available tool to conduct simulations using PALPS although in \cite{palps2} we can find the encoding and the model checking of some examples using the probabilistic model checker PRISM \cite{prism}. The translation from PALPS to PRISM supports to the study of temporal properties, but since spatial features are not explicitly represented in PRISM features, there is no straightforward or convenient structure to define and analyse spatial properties.\\
PALOMA (Process Algebra for Located Markovian Agents) \cite{paloma} is a stochastic process algebra developed to allow the expression of models representing systems composed by populations of agents distributed over space, where the relative positions of agents influence their interaction. Even if this language is not specifically created to be an ecological support, we find interesting aspects from our point of view, such as the location attributes. However, PALOMA is a conservative language, meaning agents cannot be created or destroyed, making difficult to capture the birth/death process. Moreover, PALOMA does not currently present any analysis techniques to explicitly study the spatio-temporal dynamics of the system.\\
Starting from all these important aspects, in the following sections we present our language MELA, demonstrating that it takes into account the features that have been highlighted and that it is tailored to ecological modelling.
In the literature we can find other existing formalisms, which also support explicit space, applied to ecology, like Membrane P Systems \cite{sys1}, and Cellular Automata \cite{cell}. They have inspired a new formalism, the Grid System. This has been applied, for example, to study population model of a species of mosquitos, considering spatiality, \cite{sys2} and to stochastically model dynamics of seasonal migration, where space is abstracted using a grid of cells \cite{sys3}.

\section{MELA Syntax}

In this section we present the syntax of MELA. We initially describe the component $P(l)$ that represents the agent $P$ with a location attribute $l$: in MELA, agents can make actions, change their state, be influenced by the current state of the system, including the current location of the agents, and the environment. The environment is represented as a collection of different environmental conditions $E$, captured by the component $Env$. Within the MELA syntax we distinguish \textit{atomic components}, $\mathcal{C}_{at}$, and  \textit{compound components}, $\mathcal{C}_{com}$. The atomic components $P(l), E$  represent the behaviour of a single agent, or a single environmental condition. Compound component $Com$ and $Env$ represent the parallel composition of one or more agents, or one or more or no environmental conditions, respectively.
\begin{eqnarray*}
P(l)&::=&(\alpha, r)\star P(l)\;|\;\rightarrow\{L\}(\alpha, r)\star P(l)\;| \leftarrow(\alpha, p)\star P(l) \;|\;\\
&& P(l) + Q(l)\;|\ C(l) \;|\ \mbox{\textbf{nil}} \\
E &::=& \rightarrow\{L\}(\alpha,r).E\\
Com &::=&P(l) \;|\; Com \parallel P(l) \\
Env &::=& E\;|\; Env\parallel E \;|\; \mbox{\textbf{nil}}\\
Init &::=& Com \parallel Env
\end{eqnarray*}

As stated previously, $P$ represents an agent, with location attribute $l$,  and $E$ a single environment factor (e.g.\ presence of water), $Env$, a collection of $E$, is the total environment in effect over the system,  and $Init$  denotes the initial configuration. 
 
 $(\alpha, r)\star P(l)$, where $\star ::=\; .\;|\;\uparrow | \downarrow $ and $l::=l\;|\; new(l)$, 
denotes a \textit{no-influence action}. The agent performs the action $\alpha$ with a rate $r$, representing the parameter of an exponential distribution, behaving subsequently as the agent $P$ in location $l$ ($l \in \mathcal{L}$, the set of all locations). An action $\alpha$ can create ($\uparrow$) or destroy ($\downarrow$) agents, can change the state of the agent and/or its location.  The location of the resulting agent is expressed either as a single location or as a distribution over a set of possible locations, specified by $new(l): \mathcal{L} \rightarrow Pr(N_l)$. $new(l)$
assigns a probability to the set $N_l$ ($N_l \subseteq \mathcal{L})$, the set of the possible locations in the defined neighbourhood of $l$. We can either use a given probability distribution over the neighbouring locations (e.g.\ uniform distribution $U(l_1, l_2, \cdots, l_n)$) or define a distribution empirically (e.g.\ $(l_1[p_1], l_2[p_2], \cdots, l_n[p_n])$, where $p_i$ may depend on functions  $f(\mathcal{P})$, where $f : \mathcal{P} \mapsto \mathbb{R}_{\geq 0}$ is a function over the current state of the system $\mathcal{P}$, for example capturing a dependency on the number of the agents in some specific locations $[P(l)]$.
In each case the time needed to perform the action is exponentially distributed\footnote{A random variable $T$ has \textit{exponential distribution} with parameter $\lambda$ $(0 \leq \lambda < \infty)$ if $P(T > t) = e^{- \lambda t}$, for all $t \geq 0$. The mean, or expected value, of an exponentially distributed random variable T with rate parameter $\lambda$ is given by $E[T] = \frac{1}{\lambda}$.}. Thus we assume the actions of agents have a random duration. We can also describe the rate as  a function $f(\mathcal{P})$.  

$\rightarrow\{L\}(\alpha, r)\star P(l)$,
where $L ::= l\;| \;L, l\;| \;all$ is an  \textit{influence action} that may affect agents present in the set of locations specified by the set $L$, $L \subseteq \mathcal{L}$.  This set may be a single location $l$, a list of locations or all the locations, denoted by $all$. $\leftarrow(\alpha, p)\star P(l)$ denotes the potential effect of action $\alpha$ on the agent with probability $p$ of being affected. Also in this case, the probability $p$ can be a constant value or a function of the current state of the system $f(\mathcal{P})$.

The \textit{choice operator} $+$, that represents a component which may behave either as $P(l)$
or as $Q(l)$, the \textit{constant} $C$ and the null component \mbox{\textbf{nil}} are standard. Constants, whose meaning is given by a defining equation $C(l) \stackrel[]{def}{=} P(l)$, allow us to assign names to components, giving to $C(l)$ the behaviour of component $P(l)$. This supports mutually recursion definitions, supporting the specification of recurrent behaviours over a finite set of states.

$E ::= \rightarrow\{L\}(\alpha,r).E$, denotes an environmental factor or condition.  This captures an influence which can affect the evolution of the system but is not associated with a specific agent. This factor may be in effect in multiple locations, as specified by the set $\mathcal{L}$, but the component itself has no location. 

$Com ::= P(l) \;|\; Com \parallel P(l)$ denotes a composition of agents, consisting of one or more agents in parallel.  Similarly $Env ::=  E\;|\;Env\parallel E \;|\; \mathbf{nil}$, denotes no, one or more environmental conditions that comprise the \textit{environment} in effect in the system.  We assume that the agents $E$, when they exist, behave independently, and remain unchanged throughout the evolution of the system.

$Init ::= Com \parallel Env$, defines the \textit{initial configuration} of the system, consisting of the list of components present in the initial state of the system.  Maintaining the separation between the agents and the environment means that we can readily experiment with the same composition of agents under different sets of environmental conditions.

\section{MELA Semantics} 

Before introducing the semantics, we  define a structural congruence , $\equiv$, that will allow us to define the semantics and the models in a more compact and straightforward way.
\begin{eqnarray*}
P(l)[n] &\equiv & \; \stackrel[n\mbox{ }times]{}{P(l)\parallel \cdots\parallel P(l)}\\
P(l) & \equiv & P(l)[1]\\ 
P(l)[n]\parallel \mbox{\textbf{nil}} & \equiv & P(l)[n]\\ 
P(l)[x]\parallel P(l)[y] & \equiv & P(l)[x+y] \\
P(l)\parallel Q(l) &\equiv & Q(l) \parallel P(l)\\ 
(P(l)\parallel Q(l)) \parallel R(l)& \equiv & P(l) \parallel (Q(l) \parallel R(l)) \\ 
Com_1 \parallel Com_2 & \equiv & Com_2 \parallel Com_1
\end{eqnarray*}
MELA is equipped with a \textit{structural operational semantics} \cite{sos}, which describes the possible evolutions of the components, in terms of a \textit{Labelled Transition System} (LTS) $(\mathcal{C}, \rightarrow, \Theta)$.  Here $\mathcal{C}$ is the set of components, $\rightarrow$ is the transition relation between components $(\rightarrow \; \subseteq \mathcal{C} \times \Theta \times \mathcal{C})$, and $\Theta$ is the set of labels $\mathit{(mode, influence, action, value, location)}$, containing qualitative and quantitative information about the actions. Two states $S_1$ and $S_2$ are related if there is an action that may change the state of the system from $S_1$ to $S_2$. Each label is a tuple $\mathit{(mode, influence, action, value, location)}$ where
\begin{itemize}
\item $\mathit{mode} ::= . \;| \uparrow | \downarrow | \uparrow \uparrow| \uparrow \downarrow | \downarrow \uparrow | \downarrow \downarrow $
\item $\mathit{influence}::= \emptyset \;|\; L\; |\; \leftarrow$
\item $\mathit{action} ::= $ action\_name
\item $\mathit{value} ::= r \;|\; p$
\item $location ::= l_{atomic} \;|\; -$
\end{itemize}
The first entry $\mathit{mode}$ records the effect of the action on the number of agents present in the system (. no change, $\uparrow$ increase, $\downarrow$ decrease). In the presence of an influence action, which is a synchronisation between two agents, we see two different arrows, representing the effect on the populations of influencing and influenced agents respectively. We currently restrict to binary interactions so that only one agent may be influenced in each step of the evolution; this may be expanded in future work. 
%
The second entry $\mathit{influence}$ defines the role of the agent in the action; this may be to be involved in no influence with other agents  ($\emptyset$), to influence agents in some locations (set $L$) or to be influenced by the action ($\leftarrow$). The entry $\mathit{action}$ provides the name of the action whilst the entry $\mathit{value}$ is the $\mathit{rate}$ parameter of the action, or probability of influence when $\mathit{influence}$ is equal to $\leftarrow$.  The last entry \textit{location} captures the location of the atomic component that initiates or is influenced by the action. The entry is equal to $-$ for an influence action made by an environment factor $E$, since $E$ has no location and we focus on the defined influence set $L$. 

Our rules are split into those defined for atomic components and those governing the behaviour of compound components.  We first present the atomic rules.
The first three axioms present the behaviour of the no-influence prefix terms which may give rise to  the change of state of an agent, the creation of a new agent ($\uparrow$) or the destruction of a present one ($\downarrow$). Although we could have described the creation and destruction of an agent using \textbf{nil} and the parallel composition, we adopted the designing choice of $\uparrow$ and $\downarrow$ to make the syntax more suitable and effectively understandable for ecologists. In the overall view, these latter represent respectively the increase and the decrease of the number of agents present in the system. We define a function $loc: \mathcal{C}_{at} \rightarrow \mathcal{L}$,  that returns the current location of an atomic component.\\

\begin{footnotesize}
\begin{minipage}{\linewidth}
\medskip
\AxiomC{$ $}
\AxiomC{$ $}
\LeftLabel{\sf{\textbf{no-influence action .}}\quad}
\RightLabel{\hspace*{10pt} where $\bar{l}= loc((\alpha,r).P(l))$}
\BinaryInfC{$(\alpha,r).P(l){\xrightarrow{(.,\emptyset,\alpha,r,\bar{l})}}{}P(l)$}
\DisplayProof
\end{minipage}\\

\bigskip
\noindent

\begin{minipage}{\linewidth}
\AxiomC{$ $}
\AxiomC{$ $}
\LeftLabel{\sf{\textbf{no-influence action $\downarrow$}}\quad}
\RightLabel{\hspace*{10pt} where $\bar{l}= loc((\alpha,r)\downarrow P(l))$}
\BinaryInfC{$(\alpha,r)\downarrow P(l){\xrightarrow{(\downarrow,\emptyset,\alpha,r,\bar{l})}}{}\textbf{nil}$}
\DisplayProof
\end{minipage}\\

\bigskip
\noindent

\begin{minipage}{\linewidth}
\AxiomC{$ $}
\AxiomC{$ $}
\LeftLabel{\sf{\textbf{no-influence action $\uparrow$}}\quad}
\RightLabel{\hspace*{10pt} where $\bar{l}= loc((\alpha,r)\uparrow P(l))$}
\BinaryInfC{$(\alpha,r)\uparrow P(l){\xrightarrow{(\uparrow,\emptyset,\alpha,r,\bar{l})}}{}P(l)\parallel P(l)$}
\DisplayProof
\medskip
\end{minipage}
\end{footnotesize}\\

We represent the semantic rule of the probabilistic movement as follows:\\

\begin{footnotesize}
\begin{minipage}{\linewidth}
\medskip
\AxiomC{$ $}
\AxiomC{$ $}
\LeftLabel{\sf{\textbf{probabilistic movement}}\quad}
\BinaryInfC{$(move,r).P(new(l){\xrightarrow{(.,\emptyset,\alpha,r \times \bar{p_i}, \bar{l})}}{}P(l_i)$}
\DisplayProof
\end{minipage}
\end{footnotesize}\\

where $\bar{l}= loc((move,r).P(new(l)))$ and $\begin{displaystyle} \bar{p_i}=\begin{cases} 
p_i, & \mbox{if } new(l)=(l_1[p_1], \dots, l_n[p_n]) \\ 
\frac{1}{n}, & \mbox{if } new(l)=U(l_1, \dots, l_n). 
\end{cases}
\end{displaystyle}
\medskip $ \\

We also consider the base cases for influence actions:\\

\begin{footnotesize}
\begin{minipage}{\linewidth}
\medskip
\AxiomC{$ $}
\AxiomC{$ $}
\LeftLabel{\sf{\textbf{influence action}}\quad}
\RightLabel{\hspace*{10pt} where $\bar{l}= loc(\rightarrow\{L\}(\alpha,r).P(l))$}
\BinaryInfC{$\rightarrow\{L\}(\alpha,r).P(l){\xrightarrow{(., L,\alpha,r,\bar{l})}}{}P(l)$}
\DisplayProof
\end{minipage}\\

\bigskip
\noindent

\begin{minipage}{\linewidth}
\AxiomC{$ $}
\AxiomC{$ $}
\LeftLabel{\sf{\textbf{environment factor}}\quad}
\BinaryInfC{$\rightarrow\{L\}(\alpha,r).E {\xrightarrow{(.,L,\alpha,r,-)}}{}E$}
\DisplayProof
\end{minipage}\\

\bigskip
\noindent

\begin{minipage}{\linewidth}
\AxiomC{$ $}
\AxiomC{$ $}
\LeftLabel{\sf{\textbf{passive influence action}}\quad}
\RightLabel{\hspace*{10pt} where $\bar{l}= loc(\leftarrow(\alpha,p).P(l))$}
\BinaryInfC{$\leftarrow(\alpha,p).P(l){\xrightarrow{(., \leftarrow,\alpha,p,\bar{l})}}{}P(l)$}
\DisplayProof
\medskip
\end{minipage}
\end{footnotesize}\\

We consider also the case of the increase ($\uparrow$) and decrease ($\downarrow$) of the number of agents, in presence of an influence action ($L \neq \emptyset$).\\

\begin{footnotesize}
\begin{minipage}{\linewidth}
\medskip
\AxiomC{$ $}
\AxiomC{$ $}
\LeftLabel{\sf{\textbf{influence $\downarrow$}}\quad}
\RightLabel{\hspace*{10pt} where $\bar{l}=loc(\rightarrow\{L\}(\alpha,r)\downarrow P(l))$}
\BinaryInfC{$\rightarrow\{L\}(\alpha,r)\downarrow P(l) {\xrightarrow{ {(\downarrow,L,\alpha,r,\bar{l})}}}{}\textbf{nil}$}
\DisplayProof
\end{minipage}\\

\bigskip
\noindent

\begin{minipage}{\linewidth}
\AxiomC{$ $}
\AxiomC{$ $}
\LeftLabel{\sf{\textbf{influence $\uparrow$}}\quad}
\RightLabel{\hspace*{10pt} where $\bar{l}=loc(\rightarrow\{L\}(\alpha,r)\uparrow P(l))$ }
\BinaryInfC{$\rightarrow\{L\}(\alpha,r).P(l){\xrightarrow{(\uparrow,L,\alpha,r,\bar{l})}}{}P(l)\parallel P(l)$}
\DisplayProof
\medskip
\end{minipage}
\end{footnotesize}\\

The same cases hold for a passive influence action.\\

\begin{footnotesize}
\begin{minipage}{\linewidth}
\medskip
\AxiomC{$ $}
\AxiomC{$ $}
\LeftLabel{\sf{\textbf{passive influence $\downarrow$}}\quad}
\RightLabel{\hspace*{10pt} where $\bar{l}= loc(\leftarrow(\alpha,p)\downarrow P(l))$ }
\BinaryInfC{$\leftarrow(\alpha,p)\downarrow P(l) {\xrightarrow{ {(\downarrow,L,\alpha, p,\bar{l})}}}{}\textbf{nil}$}
\DisplayProof
\end{minipage}\\

\bigskip
\noindent

\begin{minipage}{\linewidth}
\AxiomC{$ $}
\AxiomC{$ $}
\LeftLabel{\sf{\textbf{passive influence $\uparrow$}}\quad}
\RightLabel{\hspace*{10pt} where $\bar{l}=loc(\leftarrow(\alpha,p)\uparrow P(l))$}
\BinaryInfC{$\leftarrow(\alpha,p)\uparrow P(l){\xrightarrow{(\uparrow,L,\alpha, p,\bar{l})}}{}P(l)\parallel P(l)$}
\DisplayProof
\medskip
\end{minipage}
\end{footnotesize}\\

In the following standard atomic rules we use the generic label $(m,i,\alpha,v,l)$, to represent the tuple $\mathit{(mode, influence, action, value, location)}$ arising from the considered atomic component.\\
\begin{footnotesize}

\begin{minipage}{0.5\linewidth}
\medskip
\AxiomC{$P(l) \xrightarrow{(m,i,\alpha,v,l)} P'(l')$}
\LeftLabel{\sf{\textbf{choice1}}\,}
\UnaryInfC{$P(l) + Q(l) \xrightarrow{(m,i,\alpha,v,l)} P'(l')$}
\DisplayProof
\end{minipage}
\begin{minipage}{0.5\linewidth}
\AxiomC{$Q(l) \xrightarrow{(m,i,\alpha,v,l)} Q'(l')$}
\LeftLabel{\sf{\textbf{choice2}}\,}
\UnaryInfC{$P(l) + Q(l) \xrightarrow{(m,i,\alpha,v,l)}Q'(l')$}
\DisplayProof
\end{minipage}\\

\bigskip
\noindent

\begin{minipage}{\linewidth}
\AxiomC{$P(l) \xrightarrow{(m,i,\alpha,v,l)} P'(l')$}
\LeftLabel{\sf{\textbf{constant}}\quad}
\RightLabel{with $C(l) \stackrel[]{def}{=} P(l)$}
\UnaryInfC{$C(l) \xrightarrow{(m,i,\alpha,v,l)} P'(l')$}
\DisplayProof
\medskip
\end{minipage}\\

\end{footnotesize}

The following rules describe the behaviour of the atomic components with an \textit{influence} action, depending on the current location of the agents and the influence set $L$. This action may change the state of the agents ($P'$, $Q'$), their location ($l_p'$, $l_q$) or the number of agents present in the system ($\uparrow$, $\downarrow$). \\

\begin{footnotesize}
\begin{minipage}{\linewidth}
\medskip
\AxiomC{$P(l_p) \xrightarrow{(\uparrow,L,\alpha,r,-)} P(l_p)\parallel P(l_p)$}
\AxiomC{$Q(l_q) \xrightarrow{(\uparrow,\leftarrow,\alpha,p,l_q)} Q(l_q)\parallel Q(l_q)$}
\LeftLabel{\sf{\textbf{influence action1 $\uparrow$ $\uparrow$}}\quad}
\RightLabel{\hspace*{10pt} if $l_q \in L$}
\BinaryInfC{$P(l_p)\parallel Q(l_q) \xrightarrow{(\uparrow \uparrow,L,\alpha,r \times p,l_q)} P(l_p)\parallel P(l_p)\parallel Q(l_q)\parallel Q(l_q)$}
\DisplayProof
\end{minipage}\\

\bigskip
\noindent

\begin{minipage}{\linewidth}
\AxiomC{$P(l_p) \xrightarrow{(\uparrow,L,\alpha,r,-)} P(l_p)\parallel P(l_p)$}
\AxiomC{$Q(l_q) \xrightarrow{(\downarrow,\leftarrow,\alpha,p,l_q)} \textbf{nil}$}
\LeftLabel{\sf{\textbf{influence action1 $\uparrow$ $\downarrow$}}\quad}
\RightLabel{\hspace*{10pt} if $l_q \in L$}
\BinaryInfC{$P(l_p)\parallel Q(l_q) \xrightarrow{(\uparrow \downarrow,L,\alpha,r \times p,l_q)} P(l_p)\parallel P(l_p)$}
\DisplayProof
\end{minipage}\\

\bigskip
\noindent

\begin{minipage}{\linewidth}
\AxiomC{$P(l_p) \xrightarrow{(\downarrow,L,\alpha,r,-)} \textbf{nil}$}
\AxiomC{$Q(l_q) \xrightarrow{(\uparrow,\leftarrow,\alpha,p,l_q)} Q(l_q)\parallel Q(l_q)$}
\LeftLabel{\sf{\textbf{influence action1 $\downarrow$  $\uparrow$}}\quad}
\RightLabel{\hspace*{10pt} if $l_q \in L$}
\BinaryInfC{$P(l_p)\parallel Q(l_q) \xrightarrow{(\downarrow \uparrow,L,\alpha,r \times p,l_q)} Q(l_q)\parallel Q(l_q)$}
\DisplayProof
\end{minipage}\\

\bigskip
\noindent

\begin{minipage}{\linewidth}
\AxiomC{$P(l_p) \xrightarrow{(\downarrow,L,\alpha,r,-)} \textbf{nil}$}
\AxiomC{$Q(l_q) \xrightarrow{(\downarrow,\leftarrow,\alpha,p,l_q)} \textbf{nil}$}
\LeftLabel{\sf{\textbf{influence action1 $\downarrow$  $\downarrow$}}\quad}
\RightLabel{\hspace*{10pt} if $l_q \in L$}
\BinaryInfC{$P(l_p)\parallel Q(l_q) \xrightarrow{(\downarrow \downarrow,L,\alpha,r \times p,l_q)} \textbf{nil}$}
\DisplayProof
\medskip
\end{minipage}
\end{footnotesize}\\

The following rules lift the behaviour of atomic components to the level of compound components.\\

\begin{footnotesize}
\begin{minipage}{0.5\linewidth}
\medskip
\AxiomC{$P(l_p) \xrightarrow{(m,i,\alpha,v,l_p)} P'(l_p')$}
\LeftLabel{\sf{\textbf{parallel}}\,}
\UnaryInfC{$P(l_p)\parallel Com \xrightarrow{(m,i,\alpha,v,l_p)} P'(l_p')\parallel Com $}
\DisplayProof
\end{minipage}
\begin{minipage}{0.5\linewidth}
\AxiomC{$Com  \xrightarrow{(m,i,\alpha,v,l_p)} Com'$}
\LeftLabel{\sf{\textbf{parallel - Env}}\,}
\UnaryInfC{$Com \parallel Env \xrightarrow{(m,i,\alpha,v,l_p)}Com'\parallel Env$}
\DisplayProof
\end{minipage}\\

\bigskip
\noindent

\begin{minipage}{\linewidth}
\AxiomC{$E_i \xrightarrow{(.,L,\alpha,r,-)}  E_i$}
\LeftLabel{\sf{\textbf{environment action}}\quad}
\RightLabel{where $Env \!= \!E_1 \!\parallel \! \cdots \! \parallel \! E_n$ and $ 1 \leq i \leq n$ }
\RightLabel{\hspace*{10pt} where $Env \!= \!E_1 \!\parallel \! \cdots \! \parallel \! E_n$ and $ 1 \leq i \leq n$}
\UnaryInfC{$Env \xrightarrow{(.,L,\alpha,r,-)} Env$}
\DisplayProof
\medskip
\end{minipage}
\end{footnotesize}\\

The following rule represents the absence of influence of the action $\alpha$ on $Q(l_q)$. Depending on the \textit{mode} $m_p$, $P(l_p)$ will evolve in $Com_p$ while the component $Q(l_q)$ will not be updated since the influence action is not effective.

\begin{footnotesize} 
\begin{minipage}{\linewidth}
\medskip
\AxiomC{$P(l_p) \xrightarrow{(m_p,L,\alpha,r,l_p)} Com_p$}
\AxiomC{$Q(l_q) \xrightarrow{(m_q,\leftarrow,\alpha,p,l_q)} Com_q$}
\LeftLabel{\sf{\textbf{influence action - no effect}}\quad}
\RightLabel{\hspace*{10pt} if $l_q \in L$}
\BinaryInfC{$P(l_p)\parallel Q(l_q)  \xrightarrow{(m_p,L,\alpha,r \times (1-p),l_q)} Com_p \parallel Q(l_q)$}
\DisplayProof
\medskip
\end{minipage}
\end{footnotesize} 

The final set of rules capture the possible effect of influence actions at the level of compound components, considering both when an influence action is effective and when not.
In the first following rule both the atomic component are affected by the action $\alpha$ and they might be in different locations. In the conclusion we indicate $l_q$ as the \textit{location} entry, giving priority to the location where the action is felt.\\

\begin{footnotesize} 
\begin{minipage}{\linewidth}
\medskip
\sf{\textbf{influence action - compound component}}\\

\AxiomC{$Com_1 \xrightarrow{(., L,\alpha,r,l_p)} Com_1'$}
\AxiomC{$Com_2  \xrightarrow{(.,\leftarrow,\alpha,p,l_q)} Com_2'$}
\RightLabel{\hspace*{10pt} if $l_q \in L$}
\BinaryInfC{$Com_1\parallel Com_2 \xrightarrow{(.,L,\alpha,r \times p,l_q)} Com_1'\parallel Com_2'$}
\DisplayProof
\end{minipage}\\

\bigskip
\noindent

\begin{minipage}{\linewidth}
\sf{\textbf{influence action - compound component - no update}}\\

\AxiomC{$Com_1 \xrightarrow{(., L,\alpha,r,l_p)} Com_1'$}
\AxiomC{$Com_2  \xrightarrow{(.,\leftarrow,\alpha,p,l_q)} Com_2'$}
\RightLabel{\hspace*{10pt} if $l_q \in L$}
\BinaryInfC{$Com_1\parallel Com_2 \xrightarrow{(.,L,\alpha,r \times (1-p),l_p)} Com_1'\parallel Com_2$}
\DisplayProof
\end{minipage}\\

\bigskip
\noindent

\begin{minipage}{\linewidth}
\sf{\textbf{influence action - compound component - no effect}}\\

\AxiomC{$Com_1 \xrightarrow{(., L,\alpha,r,l_p)} Com_1'$}
\AxiomC{$Com_2  \xrightarrow{(.,\leftarrow,\alpha,p,l_q)} Com_2'$}
\RightLabel{\hspace*{10pt} if $l_q \notin L$}
\BinaryInfC{$Com_1\parallel Com_2 \xrightarrow{(.,L,\alpha,r,l_p)} Com_1'\parallel Com_2$}
\DisplayProof
\end{minipage}\\

\bigskip
\noindent

\begin{minipage}{\linewidth}
\sf{\textbf{influence action - Env}}\\

\AxiomC{$Env \xrightarrow{(., L,\alpha,r,-)} Env$}
\AxiomC{$Com  \xrightarrow{(.,\leftarrow,\alpha,p,l_q)} Com'$}
\RightLabel{\hspace*{10pt} if $l_q \in L$}
\BinaryInfC{$Env\parallel Com \xrightarrow{(.,L,\alpha,r \times p,l_q)} Env\parallel Com'$}
\DisplayProof
\end{minipage}\\

\bigskip
\noindent

\begin{minipage}{\linewidth}
\sf{\textbf{influence action - Env - no update}}\\

\AxiomC{$Env \xrightarrow{(., L,\alpha,r,-)} Env$}
\AxiomC{$Com  \xrightarrow{(.,\leftarrow,\alpha,p,l_q)} Com'$}
\RightLabel{\hspace*{10pt} if $l_q \in L$}
\BinaryInfC{$Env\parallel Com \xrightarrow{(.,L,\alpha,r \times (1-p),-)} Env\parallel Com$}
\DisplayProof
\end{minipage}\\

\bigskip
\noindent

\begin{minipage}{\linewidth}
\sf{\textbf{influence action - Env - no effect}}\\

\AxiomC{$Env \xrightarrow{(., L,\alpha,r,-)} Env$}
\AxiomC{$Com  \xrightarrow{(.,\leftarrow,\alpha,p,l_q)} Com'$}
\RightLabel{\hspace*{10pt} if $l_q \notin L$}
\BinaryInfC{$Env\parallel Com \xrightarrow{(.,L,\alpha,r,-)} Env\parallel Com$}
\DisplayProof
\end{minipage}\\
\end{footnotesize} 
\section{MELA Examples}
In this section we present four illustrative examples, described using MELA. We show these first applications to familiarise the reader with the language. We demonstrate its expressiveness and flexibility, handling different aspects we already introduced, such as the space abstraction and the role of the environment.
We present an SI model in a 2-cell grid, with influence actions and simple spatial description. Secondly, we present a model of prey-predator dynamics on a graph to show the range of spatial descriptions available in MELA\@. 
We continue with a model of the spread of a Cholera epidemic, presenting the environment component $Env$ and its influence action, and conclude with a model showing the spread of disease in a system with nested spatial structure, distinguishing between local and overall interactions. For each model, after an initial description, we present a table with the parameters and the set of locations needed. Afterwards, we present the components and their behaviours, followed by the initial configuration of the system. For each model we will underline important and specific aspects.
\subsection{MELA SI model with grid space description}
Our first example is a classic disease spread model consisting solely of susceptible (S) and infective (I) agents. The susceptibles are individuals who do not currently have the disease but who are susceptible to infection while infectives are individuals who have the disease and may actively spread it by passing it on to susceptible individuals through so-called contact. The model represents demography through local births and deaths and between location movement of agents. 
\begin{center}
    \begin{footnotesize} 
   \begin{tabular}{| c| p{5cm}|c p{5cm}|}
      \cline{1-4}
      \multicolumn{1}{|c|}{Parameter}  &  \multicolumn{1}{l|}{Description} &  \multicolumn{1}{c|}{} & \\ [5pt]
     \hline
       $b$, $d_S$ & birth and death rate of S agents & \multicolumn{1}{c|}{$d_I$} & death rate of I agents \\ [3pt]
       \hline  
       $m_S$  & movement rate of S agents & \multicolumn{1}{c|}{$m_I$} & movement rate of I agents \\ [3pt]
        \hline
      $p$  & probability of contact for S agents & \multicolumn{1}{c|}{ $c$} & contact rate  \\ [3pt]
       \hline \hline
       \multicolumn{1}{|c|}{ $\mathcal{L} = \{1,2\}$ } &  \multicolumn{1}{c}{set of locations} & & \\
       \hline
   \end{tabular}
       \end{footnotesize} 
       \end{center} 
The MELA expression of the model is as follows:
\begin{eqnarray*}
S(l):&=&(birth,b)\uparrow S(l) + (deathS, d_S)\downarrow S(l)+\\
     && (moveS, m_S).S(new(l)) + \leftarrow(contact, p).I(l)\\
I(l):&=& (deathI,d_I) \downarrow I(l)+(moveI, m_I).I(new(l))+\\
        &&\rightarrow\{l\}(contact,c).I(l)
\end{eqnarray*}
In this case there are no environmental factors and the initial state can be written as:
\[
\mathit{Init} = S(1)[2]\parallel S(2)[1] \parallel I(1)[1]
\]
We can represent the initial configuration as follows:
\begin{center}
\begin{tikzpicture}
\draw[xstep=1cm,ystep=1,color=gray] (0,0) grid (2,1);
\matrix[matrix of nodes,
inner sep=0pt,
anchor=south west,
nodes={inner sep=0.1pt,text width=1cm,minimum height=1cm}
]{
\mbox{\hspace*{3pt}S  S   I } & \mbox{ \hspace*{4pt} S } \\
};
\end{tikzpicture}
\end{center}
In the description of $S$ agents we have \textit{birth} and \textit{deathS} processes, the movement \textit{moveS} and the \textit{contact} action. With \textit{birth} and \textit{deathS} we change the current number of $S$ individuals present in our system, increasing ($\uparrow$) or decreasing ($\downarrow$) it respectively.
We also represent the movement with the action \textit{moveS}, that updates the location depending on the function $new(l)$, defining the potential range of destinations for the moving individuals. In this case the movement is very limited and it is easy to represent.
\[
new(l)=\begin{cases} 
2, & \mbox{if } l = 1 \\ 
1, & \mbox{if } l = 2 
\end{cases}
\]
The action \textit{contact} is of interest: with $\leftarrow$ we represent an action that influences the described component, in this case the agent $S$, with probability $p$. As we can see, we find the action \textit{contact} also in the description of the agent $I$, where we define that $I$ will affect location $l$ with a specific rate. 
In the current model, this represents the fact that infection can only happen between $S$ and $I$ agents in the same location. The number of $I$ agents can decrease, due to the action \textit{deathI} and the agents $I$ can move, according to the same $new(l)$ function. 
\subsection{MELA prey-predator model with graph description}
In this example we represent a Lotka-Volterra model, to describe the dynamics of a system in which two species interact, one as a predator and the other as prey, in a number of discrete habitat locations. In this case, the locations are defined as vertices in a graph $G$ and this contrasts with the grid-based structure of the previous example. Predator-prey dynamics are a classic demonstration of the importance of spatial heterogeneity in ecology, allowing the coexistence of predator and prey \cite{huff}.
\begin{center}
\begin{footnotesize} 
    \begin{tabular}{ | c |p{7cm} |}
    \hline
     Parameter & Description\\ [5pt]\hline
   $m_{Pd}$, $bPd$, $dPd$ & predator movement, birth and death rate\\ [3pt]\hline
       $m_{Pr}$, $bPr$, $dPr$  & prey movement, birth and death rate\\  [3pt]\hline
 $e$ & eating rate \\ [3pt] \hline
  $p$ & probability of being eaten \\ [3pt] \hline \hline
$\mathcal{L}=G=(V,E)$ \hspace*{2pt} $V= \{1,2,3,4\}$  & set of locations (where edges $E$ are specified by the set \\
$ E=\{\{2,4\},\{1,3,4\},\{2,4\},\{1,3\}\}$ &  of vertices respectively reachable from each vertex) \\ \hline
    \end{tabular}
    \end{footnotesize} 
\end{center}
The MELA model is defined as follows:
\begin{eqnarray*}
Pd(v) &=& (movePd, m_{Pd}). Pd(new(v))+(birthPd, bPd) \uparrow Pd(v) + \\
        && (deathPd,dPd) \downarrow Pd(v) + \rightarrow\{v\}(eat,e)\uparrow Pd(v)\\
Pr(v) &=& (movePr, m_{Pr}). Pr(new(v))+ (birthPr, bPr) \uparrow Pr(v) +\\ &&(deathPr,dPr) \downarrow Pr(v)+\leftarrow(eat,p) \downarrow Pr(v)
\end{eqnarray*}
In this model we do not have any environmental factors; therefore, the initial configuration is:
\[
Init = Pd(1)[10]\parallel Pd(2)[5] \parallel Pr(3)[10]\parallel Pr(4)[15]
\]
We can represent the graph spatial abstraction and the initial configuration:
\begin{center}
\begin{minipage}[c]{0.4\textwidth}
 \centering
\begin{tikzpicture}[->,>=stealth',shorten >=1pt,auto,node distance=1.8cm,main node/.style={circle,draw,font=\sffamily\footnotesize}]

  \node[main node] (1) {1};
  \node[main node] (2) [below left of=1] {2};
  \node[main node] (3) [below right of=2] {3};
  \node[main node] (4) [below right of=1] {4};

  \path[every node/.style={font=\sffamily\small}]
    (1) edge node [left] {} (4)
        edge [bend right] node[left] {} (2)
    (2) edge node [right] {} (1)
        edge node {} (4)
        edge [bend right] node[left] {} (3)
    (3) edge node [right] {} (2)
        edge [bend right] node[right] {} (4)
    (4) edge node [left] {} (3)
        edge [bend right] node[right] {} (1);
\end{tikzpicture}
\end{minipage}
\begin{minipage}[c]{0.4\textwidth}
 \centering
\begin{tikzpicture}[->,>=stealth',shorten >=1pt,auto,node distance=1.5cm,main node/.style={circle,draw,font=\sffamily\footnotesize}]

  \node[main node] (1) {10 Pd};
  \node[main node] (2) [below left of=1] {5 Pd};
  \node[main node] (3) [below right of=2] {10 Pr};
  \node[main node] (4) [below right of=1] {15 Pr};

  \path[every node/.style={font=\sffamily\small}]
    (1) edge node [left] {} (4)
        edge [bend right] node[left] {} (2)
    (2) edge node [right] {} (1)
        edge node {} (4)
        edge [bend right] node[left] {} (3)
    (3) edge node [right] {} (2)
        edge [bend right] node[right] {} (4)
    (4) edge node [left] {} (3)
        edge [bend right] node[right] {} (1);
\end{tikzpicture}
\end{minipage}
\end{center}
The function $new(v)$, describing the possible movements, is defined as follows:
\[
new(v) = U(v_1, \cdots, v_n), \mbox{ } \mbox{ } v_i \in N[v]
\]
where $N[v]$ is the $v$-th element of the set E, that represents the reachable nodes starting from the node $v$. The function $new$ also defines a uniform probability distribution over the set of reachable nodes, starting from the specific $v$.
\subsection{Cholera epidemics spread with MELA}
In the following example we present a model of the spread of a Cholera epidemic: this epidemic can spread through contact between humans (secondary infection) but also by primary infections due to the presence of contaminated water \cite{cholera}. To present the key idea we will work with a $2 \times 2$ grid, where we have the presence of water just in two of the four cells, as shown. In this model we use the external factor $E$ to influence the agents present in some locations, the ones with water present. In this case the external factor is used to model the presence of water and the resulting infection process.
\begin{center}
\begin{footnotesize} 
    \begin{tabular}{ | c| p{7.5cm} |}
    \hline
     Parameter  & Description\\ [5pt]\hline
   $b$, $d_S$, $m_S$ & birth, death and movement rate of S agents \\ [3pt]\hline
      $d_I$, $m_I$ & death and movement rate of I agents \\  [3pt]\hline
  $d_R$, $m_R$ &  death and movement rate of R agents \\ [3pt] \hline
 $c$ & contact rate of I agents\\ [3pt] \hline
 $p$ & probability of infection for S agents from I \\ [3pt] \hline
  $cE$ & contact rate of E agent (the environment)\\ [3pt] \hline
   $pE$ & probability of infection for S agents from E \\ [3pt] \hline
  $rec$ & recovery rate \\ [3pt] \hline
  $rs$ & susceptibility rate \\ [3pt] \hline \hline
 $\mathcal{L}=[0,1]\times[0,1]$  & set of locations \\ [3pt] \hline
    \end{tabular}
    \end{footnotesize} 
\end{center}
In this model, in addition to susceptibles (S) and infectives (I) individuals, we also have recovered (R), individuals who have survived the disease and have temporary immunity. The representations of the agents are as follows:
\begin{center}
\begin{minipage}{0.65\textwidth}
\begin{eqnarray*}
S(x,y):&=&(birth,b)\uparrow S(x,y) + (deathS, d_S)\downarrow S(x,y)+\\
     && (moveS, m_S).S(new(x,y)) + \leftarrow(contact, p).I(x,y)+\\
     && \leftarrow(contactE, pE). I(x,y)\\
I(x,y):&=& (deathI,d_I) \downarrow I(x,y)+(moveI, m_I).I(new(x,y))+\\
        &&\rightarrow\{(x,y)\}(contact,c).I(x,y) + (recover,rec).R(x,y)\\
R(x,y):&=&(deathR, d_R) \downarrow R(x,y)+(moveR, m_R).R(new(x,y))+\\
&& (susceptible,rs).S(x,y)\\
\end{eqnarray*}
\end{minipage}
\begin{minipage}{0.3\textwidth}
\begin{center}
\begin{tikzpicture}
\draw[xstep=1cm,ystep=1,color=gray] (0,0) grid (2,2);
\matrix[matrix of nodes,
inner sep=0pt,
anchor=south west,
nodes={inner sep=0.1pt,text width=1cm,minimum height=1cm}
]{
\hspace*{12pt}- \mbox{ }(1,0) & \hspace*{12pt}- \mbox{ }(1,1)   \\
water \mbox{ }(0,0) & water \mbox{ }(0,1) \\
};
\end{tikzpicture}
\end{center}
\end{minipage}
\end{center}
The external factor capturing the influence of the contaminated water is represented as:
\[
E = \rightarrow\{(0,0),(0,1)\}(contactE, cE).E
\]
Therefore the initial configuration can be written as:
\[
Init = S(0,0)[100]\parallel I(0,0)[1]\parallel E[1]
\]
The function $new(x,y)$ is here defined as:
\[
new(x,y) = U(((x+1)\mbox{ mod}2,y), (x,(y+1)\mbox{ mod}2))
\]
representing movement on a $2 \times 2$ grid, with periodic boundaries and a Von Neumann neighbourhood.\\
We can also consider the case in which contaminated water is present in each cell of the grid.
Only the external factor needs to be modified. For this configuration we will use $all$ when we define the action range for  $E$:
\[
E = \rightarrow\{all\}(contactE, cE). E\\
\]
The compositional high-level nature of MELA allows the modeller to easily change aspects of the model such as this, while other tools may require more pervasive changes, making it harder to maintain an overall view of the structure of the system.
\subsection{MELA SI model with nested spatial structure}
In our final example we consider a system with a more complex nested structure. We develop an SI model between nodes, e.g. social groups of animal or farms, representing each as a grid and the relations between nodes using a graph.
For this reason the location attribute will be defined with 3 entries $(x,y,v)$, where $x$ and $y$ represent the cell within the node $v$.
The relations between the sites and the local representation of nodes are represented in the following figure:
\begin{center}
\includegraphics[width=0.5\linewidth]{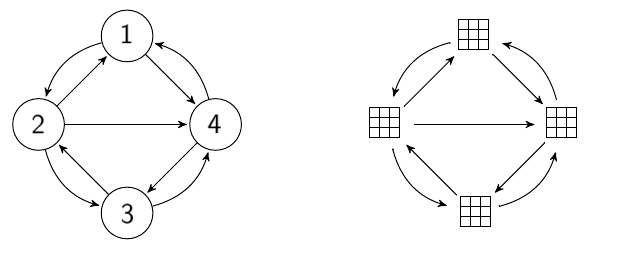}
\end{center}
The within-node structure is shown above as a $3 \times 3$ grid. Such internal structure could, for example, represent buildings and fields or multiple cattle herds within each farm or the within-group social structure of social mammals. There are two different levels of interactions, a \textit{local} set of interactions between animals in the same node (grid), and a \textit{global} relation between nodes (graph), due to the dispersal or the transport or sale of animals. For example, this model structure also accommodates the household models used to study human epidemics such as measles \cite{house}.
\begin{center}
\begin{footnotesize} 
    \begin{tabular}{ | c | p{8.8cm} |}
    \hline
     Parameter & Description\\ [5pt]\hline
   $b$, $d$& birth and death rate of S animals \\ [3pt]\hline
    $m_{iS}$,  $m_{iI}$& movement rate of S and I agents in the grid\\ [3pt] \hline
   $d_I$ & death rate of I agents \\  [3pt]\hline
  $m_{oS}$, $m_{oI}$ & movement rate of S and I agents to neighbouring nodes\\ [3pt] \hline
   $c$ & contact rate of I agents\\ [3pt] \hline
   $p$ & probability of infection for S agents from I \\ [3pt] \hline \hline
$\mathcal{L}=[0,2]\times[0,2]\times V$  & $(x,y,v) \in \mathcal{L}$, $(x,y)$: grid, $v$: graph \\ \hline
\end{tabular}
\end{footnotesize} 
\end{center}
The set $V$ is the set of vertices of the graph $G=(V,E)$, where $V= \{1,2,3,4\}$ and \\$E=\{\{2,4\},\{1,3,4\},\{2,4\},\{1,3\}\}$.\\
In this case the MELA expression of the model is as follows:
\begin{eqnarray*}
S(x,y,v):&=&(birth,b)\uparrow S(x,y,v) + (death, d)\downarrow S(x,y,v)+\\
     &&\leftarrow(contact, p).I(x,y,v)+(moveinS,m_{iS}).S(new(x,y,v))+\\
     &&(moveoutS,m_{oS}).S(new_v(x,y,v))\\
I(x,y,v):&=& (deathI,d_I) \downarrow I(x,y,v)+\rightarrow\{(x,y,v)\}(contact,c).I(x,y,v)+\\
&&(moveinI,m_{iI}).I(new(x,y,v))+(moveoutI,m_{oI}).I(new_v(x,y,v))\\
Init &=& S(0,0,1)[5] \parallel S(0,0,2)[5] \parallel S(0,0,3)[5] \parallel S(0,0,4)[5] \parallel I(1,1,1)[1]
\end{eqnarray*}
Here, the function $new$ describes the possible changes of $x$ and $y$ on a $3 \times 3$ grid, with periodic boundaries and a Von Neumann neighbourhood, leaving $v$ unchanged.
\begin{center}
$new(x,y,v) = U(((x+1)\mbox{ mod}3,y,v),(x,(y+1)\mbox{ mod}3,v), $\\
\hspace{75pt} $((x+2)\mbox{ mod}3,y,v), (x,(y+2)\mbox{ mod}3,v))$\\
\end{center}
With $new_v$ we describe the movement between nodes and we update the value $v$, choosing with the same probability between all the neighbouring vertices, defined in the set $N(v)$. We also define in which cell of the new node the animals will arrive. For example, it can be the lower left cell $(0,0)$:
\[
new_v(x,y,v) = (0,0,U(N(v)))
\]

\section{Discussion and Future Work}

We have introduced the process algebra MELA and demonstrated its application to a number of examples incorporating distinct features. We have described the semantics of MELA in terms of a labelled transition system. Starting from this we can derive the underlying continuous time Markov chain and perform stochastic simulations. We can represent the stochastic process as a family of vectors
\[
\mathbf{(P(t))}_{t \in T} = (P_1(l_1)(t), \dots, P_i(l_j)(t) ,\dots, P_m(l_n)(t))_{t \in T} 
\] 
where the vector $\mathbf{P(t)}$ represents the state of the system at time $t$ and the entries $P_i(l_j)$ count the number of agents $P_i$ in location $l_j$ at time $t$, where $i \in \{1,\dots,m\}$, $m$ is the number of different agents present in our system, $j \in \{1,\dots,n\}$ and $n$ is the number of different locations.
The dynamics of the system is specified by the transitions, that give information about the actions changing the state of the system. As is standard in stochastic process algebras, we can derive the transition rate matrix, characterising the underlying CTMC, from the labelled transition system. We can also derive a deterministic description of the system, given by a system of ODEs, approximating the model by fluid approximation. Again using the labelled transition system, we can derive a \textit{stoichiometry matrix} $M$, defining how each action changes the number of the agents in the system. Once we have defined the initial state $\mathbf{(P(0))}$ of the vector $\mathbf{(P(t))}$, the matrix $M$ and the vector $v$ of the rates, that specifies the rate function for each action, the ODE system is obtained as
\[
\frac{dP}{dt} = M \times v 
\]
This system of ODEs specifies the change on the number of agents in each different state in each location during the evolution of the system, depending on the different actions. 

Our next goal will be to implement MELA and then analyse and simulate the examples we have presented, as well as larger case studies, studying how different evolutions can arise, depending on different assumptions, such as spatial configurations, interaction rates and the role of the environment, high-level reasoning about them. Since spatial dynamics and temporal evolution of ecological systems are of key interest in our research, an interesting further step will be to develop a complementary spatio-temporal logic, suitable for formally describing and verifying spatio-temporal properties of MELA models. We will also develop a suitable statistical model checking algorithm, using MELA specific features, to facilitate being able to verify properties like ``\textit{how far the disease can spread in a specific time interval}", ``\textit{what is the probability of being infected by a neighbouring infected agent in a specific time interval}", ``\textit{what is the probability of being infected by a disease that originates from contaminated water, for agents in locations without water presence}".

\section*{Acknowledgement}

This work was supported by Microsoft Research Cambridge through its PhD Scholarship Programme. Glenn Marion wishes to acknowledge funding from the Scottish Government.

\providecommand{\urlalt}[2]{\href{#1}{#2}}
\providecommand{\doi}[1]{doi:\urlalt{http://dx.doi.org/#1}{#1}}


\begin{thebibliography}{999}
\bibitem{palps} M. Antonaki, A. Philippou, \textit{A process calculus for spatially-explicit ecological models}, Electronic Proceedings in Theoretical Computer Science (EPTCS) 100, 2012, pp. 14-28, \doi{10.4204/EPTCS.100.2}
\bibitem{house} F. Ball, \textit{Stochastic and deterministic models for SIS epidemics among a population partitioned into households}, Math Biosci. 156, 1999, pp. 41-67, \doi{10.1016/S0025-5564(98)10060-3}
\bibitem{sys2} R. Barbuti, A. Cerone, A. Maggiolo-Schettini, P. Milazzo, S. Setiawan, \textit{Modelling Population Dynamics Using Grid Systems}, Information Technology and Open Source: Applications for Education, Innovation, and Sustainability, Lecture Notes in Computer Science Volume 7991, 2014, pp. 172-189, \doi{10.1007/978-3-642-54338-8\_14}
\bibitem{pepastirling} S. Benkirane, J. Hillston, C. McCaig, R. Norman, C. Shankland, \textit{Improved Continuous Approximation of PEPA Models through Epidemiological Examples}, From Biology to Concurrency and back, FBTC 2008. Electronic Notes in Theoretical Computer Science 229(1), 2009, pp. 59-74, \doi{10.1016/j.entcs.2009.02.005}
\bibitem{cholera} E. Bertuzzo, R. Casagrandi, M. Gatto, I. Rodriguez-Iturbe, A. Rinaldo, \textit{On spatially explicit models of cholera epidemics}, J. R. Soc. Interface 7(43), 2010, pp. 321-33, \doi{10.1098/rsif.2009.0204}
\bibitem{original} F. Ciocchetta, J. Hillston, \textit{Bio-PEPA: A framework for the modelling and analysis of biological systems}, Theoretical Computer Science 410, 2009, pp. 3065-3084, \doi{10.1016/j.tcs.2009.02.037}
\bibitem{bio-pepa} F. Ciocchetta, J. Hillston, \textit{Bio-PEPA for epidemiological models}, Journal Electronic Notes in Theoretical Computer Science 261, 2010, pp. 43-69, \doi{10.1016/j.entcs.2010.01.005}
\bibitem{paloma} C. Feng, J. Hillston, \textit{PALOMA: A process algebra for located Markovian agents}, 11th International Conference on Quantitative Evaluation of Systems (QEST) 2014, pp. 265-280, \doi{10.1007/978-3-319-10696-0\_22}
\bibitem{cheng} C. Feng, J. Hillston, V. Galpin, \textit{Automatic Moment-closure Approximation of Spatially Distributed Collective Adaptive Systems}, Journal ACM Transactions on Modelling and Computer Simulation 26, 2016,  \doi{10.1145/2883608}
\bibitem{pepa} J. Hillston, \textit{A Compositional Approach to Performance Modelling}, Cambridge University Press, 1996, \doi{10.1017/cbo9780511569951}
\bibitem{huff} C. B. Huffaker, \textit{Experimental studies on predation: dispersion factors and predator-prey oscillations}, Hilgardia 27(14), 1958, \doi{10.3733/hilg.v27n14p343}
\bibitem{fmd} M. J. Keeling, \textit{Models of foot-and-mouth disease}, Proc. R. Soc. B 272, 2005, pp. 1195-1202, \doi{10.1098/rspb.2004.3046}
\bibitem{cell}  M. J. Keeling, P. Rohani, \textit{Modeling Infectious Diseases in Humans and Animals}, 2007, Princeton University Press
\bibitem{prism} M. Kwiatkowska, G. Norman, D. Parker, \textit{PRISM: Probabilistic Model Checking for Performance and Reliability Analysis}, ACM SIGMETRICS Performance Evaluation Review 36(4), 2009, pp. 40-45, \doi{10.1145/1530873.1530882}
\bibitem{carma} M. Loreti, J. Hillston, \textit{Modelling and Analysis of Collective Adaptive Systems with CARMA and its Tools}, 16th International School on Formal Methods for the Design of Computer, Communication, and Software Systems, SFM 2016,  pp. 83-119, \doi{10.1007/978-3-319-34096-8\_4}
\bibitem{foraging} G. Marion, D. Swain, M. Hutchings, \textit{Understanding foraging behaviour in spatially heterogeneous environments}, Journal of Theoretical Biology 232, 2005, pp. 127-142, \doi{10.1016/j.jtbi.2004.08.005}
\bibitem{stirling} C. McCaig, M. Begon, R. Norman, and C. Shankland, \textit{A rigorous approach to investigating common assumptions about disease transmission: Process algebra as an emerging modelling methodology for epidemiology}, Theory in Biosciences, Special Issue on Emerging Modelling Methodologies in Medicine, 2011, pp.19-29, \doi{10.1007/s12064-010-0106-8}
\bibitem{ccs} R. Milner, \textit{A Calculus of Communicating Systems}, Springer Verlag, 1980, \doi{10.1007/3-540-10235-3}
\bibitem{palps2} A. Philippou, M. Toro, M. Antonaki, \textit{Simulation and Verification for a Process Calculus for Spatially-explicit Ecological Models}, Scientific Annals of Computer Science 23(1), 2013, pp. 119-167, \doi{10.7561/SACS.2013.1.119}
\bibitem{sos} G. D. Plotkin, \textit{The Origins of Structural Operational Semantics}, J. Log. Algebr. Program. 60-61, 2004, pp. 3-15, \doi{10.1016/j.jlap.2004.03.009}
\bibitem{popred} J.C. Prentice, G. Marion, P.C.L. White, R.S. Davidson, M. R. Hutchings, \textit{Demographic Processes Drive Increases in Wildlife Disease following Population Reduction}, PLOS ONE, 2014, \doi{10.1371/journal.pone.0086563}
\bibitem{sys3} S. Setiawan, A. Cerone, \textit{Stochastic Modelling of Seasonal Migration Using Rewriting Systems with Spatiality}, Software Engineering and Formal Methods, Lecture Notes in Computer Science Volume 8368, 2014, pp. 313-328, \doi{10.1007/978-3-319-05032-4\_23}
\bibitem{spatial} D. Tilman, P. Kareiva, \textit{Spatial Ecology: The Role of Space in Population Dynamics and Interspecific Interactions}, Princeton University Press, 1998
\bibitem{wsccs} C. Tofts, \textit{Processes with probabilities, priority and time}, Formal Aspects of Computing 6, 1994, pp. 536-564, \doi{10.1007/BF01211867}
\bibitem{sys1} L. Xu, \textit{Modelling to Contain Pandemic Influenza A (H1N1) with Stochastic Membrane Systems: A Work-in-Progress Paper}, Bio-Inspired Models of Network, Information, and Computing Systems, Lecture Notes of the Institute for Computer Sciences, Social Informatics and Telecommunications Engineering Volume 87, 2012, pp. 74-81, \doi{10.1007/978-3-642-32615-8\_10}
\bibitem{primer} E. F. Zipkin, C. S. Jennelle, E. G. Cooch, \textit{A primer on the application of Markov chains to the study of wildlife disease dynamics}, Methods in Ecology and Evolution 1, 2010, pp. 192-198, \doi{10.1111/j.2041-210X.2010.00018.x}
\end{thebibliography}
\end{document}